\documentclass{article}
\usepackage{amsmath}
\usepackage{amsfonts}
\usepackage{amssymb}
\usepackage{graphicx}
\usepackage{hyperref}

\begin{document}

\title{Momentum and Uncertainty Relations in the Entropic Approach to
Quantum Theory\thanks{%
Presented at MaxEnt 2011, the 31st International Workshop on Bayesian Inference
and Maximum Entropy Methods in Science and Engineering, (July 10--15, 2011,
Waterloo, Canada) }}
\author{Shahid Nawaz\thanks{%
\texttt{sn439165@albany.edu}} \ and Ariel Caticha\thanks{%
\texttt{ariel@albany.edu}} \\
%EndAName
{\small Department of Physics, University at Albany-SUNY, }\\
{\small Albany, NY 12222, USA.}}
\date{}
\maketitle

\begin{abstract}
In the Entropic Dynamics (ED) approach to quantum theory the particles have
well-defined positions but since they follow non differentiable Brownian
trajectories they cannot be assigned an instantaneous momentum.
Nevertheless, four different notions of momentum can be usefully introduced.
We derive relations among them and the corresponding uncertainty relations.
The main conclusion is that \emph{momentum is a statistical concept}: in ED
the momenta are not properties of the particles; they are attributes of the
probability distributions.
\end{abstract}

\section{Introduction}

In the Entropic Dynamics (ED) framework quantum theory is derived as an
application of the method of maximum entropy \cite{Caticha 2011}-\cite%
{Johnson 2011}. The goal is to do for quantum mechanics what Jaynes did for
statistical mechanics \cite{Jaynes 1957}.

The basic assumption is that in addition to the particle of interest the
world contains other variables whose entropy $S$ depends on the positions $x$
of the particles, $S=S(x)$. An important new feature is that the phase of
the wave function also receives a statistical interpretation: the phase
keeps track of the entropy $S(x)$ of those extra variables.

Entropic Dynamics differs from other information-based approaches to quantum
theory in that the position observable assumes a privileged role: particles
have well-defined, albeit unknown, positions. This opens the possibility of
explaining all other observables in purely informational terms. In this
paper our specific goal is to discuss momentum.

The notion of momentum has undergone a remarkable evolution from Descartes'
early imperfect notion of a scalar \textquotedblleft quantity of
motion\textquotedblright\ to Newton's vectorial quantity of motion, then
through Lagrange's generalized momenta and Hamilton's canonical momenta to
the modern quantum version of momentum as the generator of infinitesimal
translations. Each theory of motion demands its own concept of momentum. Our
goal is to identify what concept, within the entropic framework, plays the
role of momentum.

Since particles follow Brownian trajectories that are continuous but non
differentiable it is not possible to assign an instantaneous momentum to the
particles. Nevertheless, four different notions of momentum can be usefully
introduced. They are not associated to the particles but rather to their
probability distributions: (1) the \emph{current momentum} is associated to
the velocity with which probabilities flow; (2) the \emph{drift momentum}
reflects flow along the entropy gradient; (3) the \emph{osmotic momentum} is
associated to the velocity with which probabilities diffuse; and (4) the
familiar \emph{quantum momentum} is the generator of infinitesimal
translations. We find relations among these four momenta and the
corresponding uncertainty relations. There is a formal similarity to
analogous relations derived in the context of Nelson's stochastic mechanics
\cite{Nelson 1985}-\cite{Golin 1986} and the Hall-Reginatto exact
uncertainty formalism \cite{Hall Reginatto 2002}.

We show that in the entropic framework the current momentum can reasonably
be called \emph{the} momentum: its expected value agrees with that of the
quantum momentum operator and in the classical limit it coincides with the
classical momentum. The more important conclusion, however, is that in the
entropic framework \emph{momentum is a statistical concept}. In ED, unlike
the standard interpretation of quantum mechanics, the positions of particles
have definite values just as they would in classical physics. The price we
pay for this feature is that \emph{particles do not have a momentum}. More
explicitly, momentum is not an attribute of the particles but of the
probability distributions.

Entropic dynamics is not quite a \textquotedblleft theory\textquotedblright
; rather it is a framework for the construction of theories. In addition to
familiar examples such as the standard quantum theory, its classical limit,
and the usual dissipative Brownian motion it also includes less familiar
examples. We briefly explore a new kind of model with unusual hybrid
features \cite{Caticha 2011}. The model resembles Brownian motion but there
is no energy dissipation. It obeys both the classical Hamilton-Jacobi
equation and also the usual uncertainty principle. It applies to the usual
quantum regime where $\hbar $ is not negligible but obeys a
\textquotedblleft classical\textquotedblright\ non-linear Schr\"{o}dinger
equation \cite{Holland 1993}.

Section 2 is devoted to a brief review of entropic quantum dynamics---for
details see \cite{Caticha 2011}. Momentum and uncertainty relations are
discussed in section 3, the hybrid model in section 4, and we conclude in
section 5.

\section{Entropic Quantum Dynamics}

For simplicity we discuss a single particle. The configuration space $%
\mathcal{X}$ is a flat three dimensional space with the Euclidean metric, $%
\gamma _{ab}=\delta _{ab}/\sigma ^{2}$. The full significance of the scale
factor $\sigma ^{2}$ only becomes apparent when discussing several particles
with different masses \cite{Caticha 2011}.

In addition to the particle of interest the world contains other
variables---we call them $y$. Not much needs to be known about them except
that the unknown $y$ are described by a probability distribution $p(y|x)$
that depends on the position $x$ of the particle. The entropy of the $y$
variables is given by
\begin{equation}
S[p,q]=-\int dy\,p(y|x)\log \frac{p(y|x)}{q(y)}=S(x)~.  \label{entropy s}
\end{equation}%
where $q(y)$ is some underlying measure which need not be specified further.
Since $x$ enters as a parameter in $p(y|x)$ the entropy is a function of $x$%
: $S[p,q]=S(x)$.

The probability $P(x^{\prime }|x)$ that the particle takes a short step from
$x$ to a nearby point $x^{\prime }$ is obtained using the method of maximum
entropy subject to two constraints (plus normalization). The first
constraint is that as $(x,y)$ changes to the new $(x^{\prime },y^{\prime })$
the uncertainty in the new $y^{\prime }$ depends only on the new position $%
x^{\prime }$, and not on any previous value $x$, that is, $p(y|x)$ changes
to the corresponding $p(y^{\prime }|x^{\prime })$. The second constraint
reflects the physical fact that motion is continuous, that is, motion over
large distances happens through the successive accumulation of many short
steps, $\Delta x=x^{\prime }-x$. We require that the expectation $%
\left\langle \Delta \ell ^{2}\right\rangle =\left\langle \gamma _{ab}\Delta
x^{a}\Delta x^{b}\right\rangle $ be some small numerical value, which we
take to be independent of $x$ in order to reflect the translational symmetry
of the space $\mathcal{X}$. The resulting transition probability from $x^{a}$
to $x^{\prime a}=x^{a}+\Delta x^{a}$ is \cite{Caticha 2011}%
\begin{equation}
P\left( x^{\prime }|x\right) \approx \frac{1}{Z\left( x\right) }\exp \left[ -%
\frac{\tau }{2\sigma ^{2}\Delta t}\delta _{ab}\left( \Delta x^{a}-\Delta
\overline{x}^{a}\right) \left( \Delta x^{b}-\Delta \overline{x}^{b}\right) %
\right] ~.
\end{equation}%
where $\tau $ is a constant that defines the units of the time interval $%
\Delta t$ \cite{Caticha 2010}; $\Delta \overline{x}^{a}$ is the expected
step,
\begin{equation}
\Delta \overline{x}^{a}=\left\langle \Delta x^{a}\right\rangle =b^{a}\left(
x\right) \Delta t\hspace{0.2in}\text{where}\hspace{0.2in}b^{a}\left(
x\right) =\frac{\sigma ^{2}}{\tau }\delta ^{ab}\partial _{b}S\left( x\right)
~  \label{drift}
\end{equation}%
is the drift velocity. Any displacement $\Delta x^{a}$ can be expressed as
an expected drift plus a fluctuation $\Delta w^{a}$,%
\begin{equation}
\Delta x^{a}=b^{a}\left( x\right) \Delta t+\Delta w^{a},
\end{equation}%
where%
\begin{equation}
\left\langle \Delta w^{a}\right\rangle =0\hspace{0.2in}\text{and}\hspace{%
0.2in}\left\langle \Delta w^{a}\Delta w^{b}\right\rangle =\frac{\sigma ^{2}}{%
\tau }\Delta t\delta ^{ab}~.  \label{fluc}
\end{equation}%
Since the fluctuations are the order of $\Delta w\sim (\Delta t)^{1/2}$ the
trajectory of the particle is continuous but not differentiable as in
Brownian motion.

Standard methods show that the successive iteration of $P\left( x^{\prime
}|x\right) $ yields a probability distribution $\rho \left( x,t\right) $
that evolves according to Fokker-Planck equation
\begin{equation}
\partial _{t}\rho =-\partial _{a}\left( \rho v^{a}\right)  \label{FPeq}
\end{equation}%
where $v^{a}$ is the current velocity defined by%
\begin{equation}
v^{a}=b^{a}+u^{a}  \label{current vel1}
\end{equation}%
where%
\begin{equation}
u^{a}=-\frac{\sigma ^{2}}{\tau }\partial ^{a}\log \rho ^{1/2}\text{ ,}
\label{osmotic vel1}
\end{equation}%
is the osmotic velocity. The drift velocity reflects motion up the entropy
gradient, while the osmotic velocity reflects diffusion as can be seen when
written as $\rho u^{a}\propto \partial ^{a}\rho $, which is Fick's law of
diffusion. The current velocity can also be written as%
\begin{equation}
v^{a}=\frac{\sigma ^{2}}{\tau }\partial ^{a}\phi \hspace{0.3in}\text{with}%
\hspace{0.3in}\phi \left( x,t\right) =S\left( x\right) -\log \rho
^{1/2}\left( x,t\right)  \label{current vel2}
\end{equation}

The dynamics described so far is pure diffusion. In quantum mechanics, the
wave function has two degrees of freedom, the amplitude and the phase. So
far we have only one degree of freedom and that is $\rho \left( x,t\right) .$
In order to promote $\phi \left( x,t\right) $ to a dynamical degree of
freedom we allow $p(y|x)$ and $S\left( x\right) $ to be functions of time, $%
S=S\left( x,t\right) $. The time evolution of $S\left( x,t\right) $ is
determined by imposing yet another constraint, that a certain quantity ---
an \textquotedblleft energy\textquotedblright\ ---\ be conserved. Thus we
impose that the diffusion be non-dissipative. To this end introduce an
energy functional,%
\begin{equation}
E[\rho ,S]=\int d^{3}x\rho \left( x,t\right) \left( \frac{1}{2}mv^{2}+\frac{1%
}{2}\mu u^{2}+V\left( x\right) \right)   \label{energy}
\end{equation}%
where $m$ and $\mu $ are constants that will be called the mass and the
osmotic mass respectively. Imposing that the energy be conserved for
arbitrary initial choices of $\rho $ and $S$ leads to the quantum
Hamilton-Jacobi equation,
\begin{equation}
\eta \dot{\phi}+\frac{\eta ^{2}}{2m}\left( \partial _{a}\phi \right) ^{2}+V-%
\frac{\mu \eta ^{2}}{2m^{2}}\frac{\nabla ^{2}\rho ^{1/2}}{\rho ^{1/2}}=0~,
\label{QHJ}
\end{equation}%
where we have defined a new constant $\eta $ so that $\eta \overset{\text{def%
}}{=}m\sigma ^{2}/\tau $.

Eqs.(\ref{FPeq}) and (\ref{QHJ}) can be combined into a single complex
equation by using $\Psi =\rho ^{1/2}e^{i\phi }$%
\begin{equation}
i\eta \dot{\Psi}=-\frac{\eta ^{2}}{2m}\nabla ^{2}\Psi +V\Psi +\frac{\eta ^{2}%
}{2m}\left( 1-\frac{\mu }{m}\right) \frac{\nabla ^{2}\left( \Psi \Psi ^{\ast
}\right) ^{1/2}}{\left( \Psi \Psi ^{\ast }\right) ^{1/2}}\Psi ~.
\end{equation}%
In \cite{Caticha 2011} we showed that it is possible to change units and
rescale $\eta \rightarrow \kappa \eta $ and $\tau \rightarrow \kappa \tau $
by some constant $\kappa $, while simultaneously introducing $\kappa \phi _{%
\text{new}}=\phi $ and $\mu _{\text{new}}=\kappa ^{2}\mu $. This means that
there are essentially two possibilities: all theories with $\mu >0$ are
physically equivalent in that they can be rescaled/regraduated to a theory
with $\mu _{\text{new}}=m$. Setting $\kappa \eta =\hbar $ results in the Schr%
\"{o}dinger equation,%
\begin{equation}
i\hbar \frac{\partial \Psi }{\partial t}=-\frac{\hbar ^{2}}{2m}\nabla
^{2}\Psi +V\Psi ~.\hspace{0.2in}  \label{SE}
\end{equation}%
The other possibility occurs for $\mu =0$ which allows no regraduation and
leads to a non-linear Schr\"{o}dinger equation,
\begin{equation}
i\hbar \dot{\Psi}=-\frac{\hbar ^{2}}{2m}\nabla ^{2}\Psi +V\Psi +\frac{\hbar
^{2}}{2m}\frac{\nabla ^{2}\left( \Psi \Psi ^{\ast }\right) ^{1/2}}{\left(
\Psi \Psi ^{\ast }\right) ^{1/2}}\Psi ~.  \label{NLSE}
\end{equation}%
This case will be further discussed in section 4. This concludes our brief
review.

\section{Momentum in Entropic Dynamics}

During the transition from classical to quantum mechanics a central problem
was to identify some concept that would correspond to the classical momentum
in some appropriate limit. We face an analogous (but easier) problem:\ our
goal is to identify what concept, within the entropic framework, may
reasonably be called momentum.

Since the particle follows a Brownian non-differentiable trajectory it is
clear that the classical momentum $md\vec{x}/dt$ along the trajectory cannot
be defined. The obvious momentum candidates correspond to the various
velocities available to us. Thus, we define the drift, osmotic, and current
momenta,%
\begin{eqnarray}
\vec{p}_{d} &=&m\vec{b}=\hbar \vec{\nabla}S\,,  \label{pd} \\
\vec{p}_{o} &=&m\vec{u}=-\hbar \vec{\nabla}\log \rho ^{1/2}  \label{po} \\
\vec{p}_{c} &=&m\vec{v}=\hbar \vec{\nabla}\phi ~,  \label{pc}
\end{eqnarray}%
where $\phi $ is given in eq.(\ref{current vel2}). The fourth notion of
momentum that one can introduce in ED is the differential operator that
generates infinitesimal translations---it coincides, of course, with the
standard quantum momentum $\vec{p}_{q}=-i\hbar \vec{\nabla}$. Notice that
the three momenta $\vec{p}_{d}$, $\vec{p}_{o}$, and $\vec{p}_{c}$ are local
functions of $\vec{x}$ and this makes them conceptually very different from
the momentum operator $\vec{p}_{q}$. To explore their differences and
similarities we calculate the first and second moments.

\subsection*{Expected Values}

The important theorem here is the vanishing expectation of the osmotic
momentum. Using (\ref{po}) and since $\rho $ vanishes at infinity,
\begin{equation}
\langle p_{o}^{a}\rangle =-\hbar \int d^{3}x\,\rho \,\partial ^{a}\log \rho
^{1/2}=-\frac{\hbar }{2}\int d^{3}x\,\partial ^{a}\rho =0~.  \label{exp po}
\end{equation}%
Since $p_{c}=p_{d}+p_{o}$ the immediate consequence is that $\langle
p_{c}^{a}\rangle =\langle p_{d}^{a}\rangle $.

To study the connection to the quantum mechanical momentum we calculate%
\begin{equation}
\langle p_{q}^{a}\rangle =\int d^{3}x\Psi ^{\ast }\frac{\hbar }{i}\partial
^{a}\,\Psi ~.
\end{equation}%
Using $\Psi =\rho ^{1/2}e^{i\left( S-\log \rho ^{1/2}\right) }$ and (\ref%
{exp po}) one gets%
\begin{equation}
\langle p_{q}^{a}\rangle =-i\hbar \int d^{3}x\rho \left( \partial ^{a}\log
\rho ^{1/2}+i\partial ^{a}S-i\partial ^{a}\log \rho ^{1/2}\right) =\hbar
\left\langle \partial ^{a}S\right\rangle ~.
\end{equation}%
Therefore%
\begin{equation}
\langle \vec{p}_{q}\rangle =\langle \vec{p}_{c}\rangle =\langle \vec{p}%
_{d}\rangle ~,  \label{exp p}
\end{equation}%
the expectations of quantum momentum, current momentum and drift momentum
coincide.

\subsection*{Uncertainty Relations}

We start by stating a couple of definitions and an inequality. The variance
of a quantity $A$ is
\begin{equation}
\text{Var}A=\langle \left( A-\left\langle A\right\rangle \right) ^{2}\rangle
=\left\langle A^{2}\right\rangle -\left\langle A\right\rangle ^{2},
\end{equation}%
and its covariance with $B$ is
\begin{equation}
\text{Cov}\left( A,B\right) =\left\langle \left( A-\left\langle
A\right\rangle \right) \left( B-\left\langle B\right\rangle \right)
\right\rangle =\left\langle AB\right\rangle -\left\langle A\right\rangle
\left\langle B\right\rangle ~.
\end{equation}%
The general form of uncertainty relation to be used below follows from the
Schwarz inequality,
\begin{eqnarray}
\left( \text{Var}A\right) \left( \text{Var}B\right) &=&\left\langle \left(
A-\left\langle A\right\rangle \right) ^{2}\right\rangle \left\langle \left(
B-\left\langle B\right\rangle \right) ^{2}\right\rangle  \notag \\
&\geq &\left\vert \left\langle \left( A-\left\langle A\right\rangle \right)
\left( B-\left\langle B\right\rangle \right) \right\rangle \right\vert ^{2}=%
\text{Cov}^{2}\left( A,B\right) ~.  \label{main ineq}
\end{eqnarray}%
Next we apply these notions to the various momenta. An analogous calculation
in the context of stochastic mechanics is given in \cite{Golin 1985}.

\subsubsection*{Osmotic Momentum}

For simplicity we consider the one-dimensional case. The generalization is
immediate. Eq. (\ref{main ineq}) is
\begin{equation}
\left( \text{Var }x\right) \left( \text{Var }p_{o}\right) \geq \text{Cov}%
^{2}\left( x,p_{o}\right) ~.
\end{equation}%
Using (\ref{po}) and (\ref{exp po}) we have
\begin{eqnarray}
\text{Cov}\left( x,p_{o}\right) &=&\langle xp_{o}\rangle -\langle x\rangle
\langle p_{o}\rangle =\langle xp_{o}\rangle  \notag \\
&=&-\hbar \int dx\,\rho x\partial \log \rho ^{1/2}=\frac{\hbar }{2}
\label{cov osmo}
\end{eqnarray}%
Therefore%
\begin{equation}
\left( \text{Var }x\right) \left( \text{Var }p_{o}\right) \geq (\frac{\hbar
}{2})^{2}\quad \text{or}\quad \Delta x\,\Delta p_{o}\geq \frac{\hbar }{2}~,
\label{UR osmo}
\end{equation}%
which coincides with the Heisenberg uncertainty relation.

\subsubsection*{Drift Momentum}

The uncertainty relation is
\begin{equation}
\left( \text{Var }x\right) \left( \text{Var }p_{d}\right) \geq \text{Cov}%
^{2}\left( x,p_{d}\right)
\end{equation}%
Consider
\begin{eqnarray}
\text{Cov}\left( x,p_{d}\right) &=&\langle xp_{d}\rangle -\langle x\rangle
\langle p_{d}\rangle  \notag \\
&=&\hbar \int dx\,\rho x\partial S-(\int dx\,\rho x)(\hbar \int dx\,\rho
\partial S)~.
\end{eqnarray}%
The integrands involve $\rho $ and $\partial S$ which can be chosen
independently. We can choose as narrow a probability distribution as we
like, for example $\rho \rightarrow \delta \left( x-x_{0}\right) $, which
trivially leads to Cov$\left( x,p_{d}\right) =0$. Therefore, the uncertainty
relation for drift momentum is
\begin{equation}
\left( \text{Var }x\right) \left( \text{Var }p_{d}\right) \geq \text{Cov}%
^{2}\left( x,p_{d}\right) \geq 0\quad \text{or}\quad \Delta x\,\Delta
p_{d}\geq 0~.  \label{UR drift}
\end{equation}

\subsubsection*{Quantum Momentum: the Schr\"{o}dinger and the Heisenberg
Uncertainty Relations}

There appears to be no useful insight to be found from the uncertainty
relation for current momentum. It is nevertheless true that
\begin{equation}
\left( \text{Var }x\right) \left( \text{Var }p_{c}\right) \geq \text{Cov}%
^{2}\left( x,p_{c}\right)  \label{UR curr}
\end{equation}%
Since $p_{c}=p_{d}+p_{o}$, we have%
\begin{equation}
\text{Cov}\left( x,p_{c}\right) =\text{Cov}\left( x,p_{d}\right) +\text{Cov}%
\left( x,p_{o}\right)
\end{equation}

Let us now focus our attention on the quantum momentum. Using $\Psi =\rho
^{1/2}e^{i\phi }$, (\ref{po}) and (\ref{pc}) we have, after an integration
by parts,
\begin{equation}
\langle p_{q}^{2}\rangle =\int dx\Psi ^{\ast }(\frac{\hbar }{i}\partial
)^{2}\,\Psi =\langle p_{c}^{2}\rangle +\langle p_{o}^{2}\rangle ~.
\end{equation}%
Together with eqs.(\ref{exp po}) and (\ref{exp p}) this leads to
\begin{equation}
\text{Var }p_{q}=\langle p_{q}^{2}\rangle -\langle p_{q}\rangle ^{2}=\text{%
Var }p_{c}+\text{Var }p_{o}~.
\end{equation}%
Combining inequalities (\ref{UR osmo}) and (\ref{UR curr}) gives,
\begin{equation}
\left( \text{Var }x\right) \left( \text{Var }p_{q}\right) \geq \text{Cov}%
^{2}\left( x,p_{c}\right) +(\frac{\hbar }{2})^{2}~.
\end{equation}%
Finally, a straightforward calculation gives
\begin{eqnarray}
\text{Cov}\left( x,p_{q}\right) &=&\frac{1}{2}\langle xp_{q}+p_{q}x\rangle
-\langle x\rangle \langle p_{q}\rangle  \notag \\
&=&\text{Cov}\left( x,p_{c}\right) ~.
\end{eqnarray}%
Therefore,
\begin{equation}
\left( \text{Var }x\right) \left( \text{Var }p_{q}\right) \geq \text{Cov}%
^{2}\left( x,p_{q}\right) +(\frac{\hbar }{2})^{2}~,
\end{equation}%
which is a version of the quantum uncertainty relation originally proposed
by Schr\"{o}dinger \cite{Golin 1985}. Since Cov$^{2}\left( x,p_{q}\right)
\geq 0$ the somewhat weaker Heisenberg uncertainty relation follows
immediately,
\begin{equation}
\left( \text{Var }x\right) \left( \text{Var }p_{q}\right) \geq (\frac{\hbar
}{2})^{2}\quad \text{or}\quad \Delta x\,\Delta p_{q}\geq \frac{\hbar }{2}~.
\end{equation}

\section{A Hybrid Theory}

Non-dissipative ED is defined by the Fokker-Planck equation (\ref{FPeq}) and
the quantum Hamilton-Jacobi eq.(\ref{QHJ}). Here we focus on the special
case with $\mu =0$. Setting $\hbar =\eta $ and $S_{HJ}=\hbar \phi $ in eq.(%
\ref{QHJ}) gives
\begin{equation}
\dot{S}_{HJ}+\frac{1}{2m}\left( \vec{\nabla}S_{HJ}\right) ^{2}+V=0\text{ , \
\ \ \ \ \ }  \label{HJ}
\end{equation}%
which is classical Hamilton-Jacobi equation.

One might be tempted to interpret the $\mu =0$ model as a classical ensemble
dynamics but this is wrong. To see this it is useful to contrast $\mu =0$
with the usual classical limit defined by $\hbar /m\rightarrow 0$. As $\hbar
\rightarrow 0$ with $S_{HJ}$, $m$, and $\mu $ fixed, the current and the
osmotic momenta, given in (\ref{pc}) and (\ref{po}), become
\begin{equation}
\vec{p}_{c}=m\vec{v}=\,\vec{\nabla}S_{HJ}\quad \text{and\quad }\vec{p}_{o}=m%
\vec{u}=0~,
\end{equation}%
and $S_{HJ}$ satisfies the classical eq.(\ref{HJ}). Furthermore, according
to eq.(\ref{drift}) the particle is expected to move along the entropy
gradient, while eq.(\ref{fluc}),
\begin{equation}
\left\langle \Delta w^{a}\right\rangle =0\quad \text{and}\quad \left\langle
\Delta w^{a}\Delta w^{b}\right\rangle =\frac{\hbar }{m}\Delta t\,\delta
^{ab}\rightarrow 0~,  \label{Zero Fluct}
\end{equation}%
shows that the fluctuations about the expected trajectory vanish. Therefore,
in the limit $\hbar \rightarrow 0$ ED reproduces classical mechanics with
classical trajectories following the entropy gradient. The same conclusion
is obtained for fixed $\hbar $ provided the mass $m$ is sufficiently large.

The limit $\mu \rightarrow 0$ is very different! Here $\hbar $ and $m$ are
fixed and $\hbar /m$ need not be small. This situation is also ruled by the
classical Hamilton-Jacobi equation (\ref{HJ}), but the osmotic momentum does
not vanish,
\begin{equation}
\vec{p}_{c}=m\vec{v}=\,\vec{\nabla}S_{HJ}\quad \text{and\quad }\vec{p}_{o}=m%
\vec{u}=-\hbar \vec{\nabla}\log \rho ^{1/2}~~.
\end{equation}%
The expected trajectory lies along a classical path but now the fluctuations
$\Delta w^{a}$ about the classical trajectory, eq.(\ref{Zero Fluct}) no
longer vanish.

All the considerations about momentum described in the previous section
apply to this $\mu =0$ model. In particular, the momentum operator $\vec{p}%
_{q}=-i\hbar \vec{\nabla}$ can be introduced---for exactly the same reasons
that one would introduce it in quantum theory---as a generator of
translations, and this means that the $\mu =0$ model obeys uncertainty
relations identical to quantum theory. And yet, this is not quantum theory:
the corresponding Schr\"{o}dinger equation, eq.(\ref{NLSE}), is nonlinear
and therefore there is no superposition principle.

\section{Conclusion}

We have explored the notion of momentum in entropic quantum dynamics. We
find that the current momentum can reasonably be called \emph{the} momentum
because its expected value agrees with that of the quantum momentum operator
and in the classical limit it coincides with the classical momentum. We have
derived uncertainty relations within the entropic framework. A new insight
is the reason the Heisenberg relation arises which is traced to the peculiar
form of the osmotic velocity---it is essentially a diffusion effect
described by Fick's law.

The main conclusion is that in the entropic framework \emph{momentum is a
statistical concept}. In ED, unlike the standard interpretation of quantum
mechanics, the positions of particles have definite values just as they
would in classical physics. The price we pay for this feature is that \emph{%
particles do not have a momentum}. More explicitly, momentum is not an
attribute of the particles but of the probability distributions.

Finally, we explored ED for $\mu =0$ which yields what we believe to be an
altogether new kind of theory, neither classical nor quantum---we call it a
hybrid theory. Whether it can usefully describe any actual physical system
remains to be seen.

\section*{Acknowledgments}

We would like to thank A. Inomata, D. T. Johnson, and M. Reginatto for many
valuable discussions.

\end{document}